\documentclass[prb,nofootinbib,twocolumn,superscriptaddress]{revtex4} 
\usepackage{graphicx}% Include figure files
\usepackage{dcolumn}% Align table columns on decimal point
\usepackage{bm}% bold math 
\usepackage{threeparttable}
\usepackage{times}
\usepackage{mathptmx}
\usepackage{lscape}
\usepackage{natbib}
\usepackage{amsmath}
\usepackage{amssymb}
\usepackage{braket}
\usepackage{comment}
\usepackage{color}

\def\degree{\kern-.2em\r{}\kern-.3em}

\begin{document}

%\preprint{APS/123-QED} 

\title{ A Single Microscopic State to Characterize Ordering Tendency\\ in Discrete Multicomponent System   }

\author{Koretaka Yuge}
\affiliation{
Department of Materials Science and Engineering,  Kyoto University, Sakyo, Kyoto 606-8501, Japan\\
}%

\author{Shouno Ohta}
\affiliation{
Department of Materials Science and Engineering,  Kyoto University, Sakyo, Kyoto 606-8501, Japan\\
}%

\begin{abstract}
{ %\begin{itemize}
%\item 
Our recent study reveals that macroscopic structure in thermodynamically equilibrium state and its temperature dependence for classical discrete system can be well-characterized by a single specially-selected microscopic state (which we call "projection state: PS"), whose structure can be known \textit{a priori} without any information about energy or temperature. 
%\item
Although PS can be universally constructed for any number of components $R$, practical application of PS to systems with $R\ge 3$ is non-trivial compared with $R=2$ (i.e., binary system). 
%\item
This is because (i) essentially, multicomponent system should inevitably requires linear transformation from conventional basis functions to intuitively-interpreted cluster probability basis, i.e., multiple PS energies are required to predict one chosen pair probability, leading to practically accumulating numerical errors, and (ii) additionally, explicit formula for the transformation from basis functions to pair probabilities should be required, which has been explicitly provided up to ternary ($R=3$) system so far.
%\item
%\item 
We here derive modified formulation to directly determine probability for like- and unlike-atom pair consisting of any chosen elements by using a single PS energy, with providing explicit relationship between basis functions and pair probabilities up to quinary ($R=5$) systems. We demonstrate the validity of the formulation by comparing temperature dependence of pair probabilities for multicomponent systems from thermodynamic simulations.  

%\end{itemize}
  }
\end{abstract}

%\pacs{81.90.+c \sep 61.05.-a \sep 05.20.Gg \sep 05.10.-a \sep 02.30.Zz }

\maketitle

\section{Introduction}
%\begin{itemize}
%\item 
For substitutional crystalline solids considered as classical, many-body, discrete system under constant composition, dynamical variables in thermodynamically equilibrium state can be determined from well-known thermodynamic average including information about possible microscopic states on phase space (or configuration space). The number of states exponentially increases with system size increased, making it practically intractable to directly determine macroscopic properties considering all possible states. Therefore, alternative approaches have been developed such as Metropolis algorism, entropic sampling and Wang-Landau sampling for efficient exploration of important microscopic states to determine equilibrium properties.\cite{mc1, mc2, mc3, wl}  Since internal energy $E$ depends on given many-body interaction, a set of such important microscopic state certainly depends on temperature as well as on interaction through the Boltzmann factor, $\exp\left(-\beta E\right)$.
%\item
Despite these facts, we recently reveal that macroscopic structure in equilibrium state along chosen coordination can be well-characterized by a single specially-selected microscopic state (which we call "projection state: PS"), whose structure can be known \textit{a priori} without any information about energy or temperature.\cite{em1,em2,em3}  In our previous study, although we show that PS can be in principle constructed for any number of constituents $R$, application only to binary system has been demonstrated combined with \textit{ab initio} calculation, where application to multicomponent system is practically non-trivial.
%\item
This is because in alloys with given composition, their microscopic structure (in configuration space) is practically described by generalized Ising model for alloy thermodynamics, which employs complete orthonormal basis functions in order to quantitatively express dynamical variables (such as energy and elastic modulus) as a function of atomic configuration. 
For multicomponent systems ($R\ge 3$), prediction of equilibrium structure along intuitively-interpreted coordination (e.g., like- and unlike-atom pair clusters) requires linear combination of basis functions, since pair cluster probabilities are linearly-dependent and are not orthonormal with each other, i.e., information about physical quantities for multiple PSs are required to obtain even a single chosen pair correlation. Such linear combination should lead to accumulate errors in predicted  equilibrium structures due to the use of multiple PS energy. 
To avoid this problem, explicit relationships between conventional basis functions for generalized Ising model and intuitively-interpreted pair cluster probability should firstly be clarified, which is, to our best knowledge, available only for binary and ternary system so far.\cite{ternary}
%\item
%\item
%\item
We here propose construction of PS in multicomponent generalized Ising systems at \textit{constant} composition, where macroscopic structure along like- and/or unlike-atom pair consisting of a chosen single element can be addressed by physical quantities of a single PS, without performing linear transformation of coordination.
 by deriving complete relationships between pair probabilities and conventional basis functions in quaternary and quinary systems, which is then applied to the presently-modified formulation of equilibrium macroscopic structure by employing multiple set of basis functions. The details are shown below. 

%\end{itemize}

\section{Derivation and Applications}
%\begin{itemize}
%\item 
\subsection{Pair probability for multicomponent systems}
In generalized Ising system, complete orthonormal basis functions are generally used for given lattice $L$ with $N$ lattice points, by applying Gram-Schmidt technique to linearly-independent polynomial set $\left\{1, \sigma_i , \ldots, \sigma_i^{\left(R-1\right)}\right\}$ for $R$-component system ($\sigma_i$ denotes spin variable to specify occupation of chosen element), namely
\begin{eqnarray}
\label{eq:b}
V_{L} &=& \bigotimes_{i=1}^{N} V_{i} \nonumber \\
V_i &=& \mathrm{span}\left\{\phi_0\left(\sigma_i\right),\ldots , \phi_{R-1}\left(\sigma_i\right)\right\} \nonumber \\
\phi_m\left(\sigma_i\right) &=& f_m\left(\sigma_i\right)/\Braket{f_m\left(\sigma_i\right)|f_m\left(\sigma_i\right)}^{1/2} \nonumber \\
f_m\left(\sigma_i\right)&=&\sigma_i^m -\sum_{j=0}^{m-1}\Braket{\phi_j\left(\sigma_i\right)|\sigma_i^m}\phi_j\left(\sigma_i\right) \quad \left(m\neq 0\right) \nonumber \\
f_0\left(\sigma_i\right) &=& 1. 
\end{eqnarray}
%%\item 
Here, $\bigotimes$ represents tensor product for vector space $V_p$ on lattice point $p$, and $\Braket{\cdot|\cdot}$ denotes inner product on configuration space. 
The definition for values of spin variable is typically symmetric, e.g., $\sigma_i = +1, 1$ for binary, $\sigma_i = +1, 0, -1$ for ternary, $\sigma_i = +2, +1, -1, -2$ for quaternary and $\sigma_i = +2, +1, 0, -1, -2$ for quinary system. For instance, from Eq.~(\ref{eq:b}), basis function on a single lattice point $k$ for binary system is given by
\begin{eqnarray}
\phi_0 = 1, \quad \phi_1 = \sigma_k,
\end{eqnarray}
where subscript denotes basis function index. Therefore, basis function for given figure $\alpha$ is given by
\begin{eqnarray}
\label{eq:2}
\Phi = \Braket{\prod_{i\in\alpha} \sigma_i},
\end{eqnarray}
where $\Braket{\quad}$ denotes taking linear average over symmetry-equivalent figure to $\alpha$. It is clear from Eq.~(\ref{eq:2}) that for binary system, sign of $\Phi_\alpha$ for pair figure directly denotes its cluster probability, i.e., whether like- ($\Phi > 0$) or unlike-atom ($\Phi < 0$) pair has larger number than another. 
%\item
Meanwhile, for multicomponent system, such intuitive interpretation does not hold. For instance, basis functions for ternary system at a single lattice point are given by
\begin{eqnarray}
\phi_0 = 1,\quad \phi_1 = \sqrt{\frac{3}{2}}\sigma_k,\quad \phi_2 =\sqrt{2}\left(\frac{3}{2}\sigma_k^2 -1 \right).
\end{eqnarray}
In this case, relationships between conventional basis functions and pair probability pair probability $y{ij}$ ($i,j=a,b,c$) are given by
%\begin{widetext}
\begin{eqnarray}
\Phi_{11} &=& \left(\sqrt{\frac{3}{2}}\right)^2 \sum_{i,j} y_{ij}\cdot \sigma_i\sigma_j  \nonumber \\
&=&\frac{3}{2} \left( yaa - 2yac + ycc  \right) \nonumber \\
\Phi_{12} &=& \frac{\sqrt{3}}{2} \left( yaa - 2yab +2ybc - ycc \right) \nonumber \\
\Phi_{22} &=& \frac{1}{2} \left( yaa-4yab+2yac+4ybb-4ybc+ycc \right),
\end{eqnarray}
%\end{widetext}
and thereby
%\begin{widetext}
\begin{eqnarray}
yaa &=& \frac{2}{3}c_a+\frac{1}{6}\Phi_{11}+\frac{\sqrt{3}}{9}\Phi_{12}+\frac{1}{18}\Phi_{22}-\frac{1}{9} \nonumber \\
yab &=& \frac{1}{3}c_a+\frac{1}{3}c_b-\frac{\sqrt{3}}{9}\Phi_{12}-\frac{1}{9}\Phi_{22}-\frac{1}{9} \nonumber \\ 
yac &=& -\frac{1}{3}c_b-\frac{1}{6}\Phi_{11}+\frac{1}{18}\Phi_{22}+\frac{2}{9} \nonumber \\ 
ybb &=& \frac{2}{3}c_b+\frac{2}{9}\Phi_{22}-\frac{1}{9} \nonumber \\ 
ybc &=& -\frac{1}{3}c_a+\frac{\sqrt{3}}{9}\Phi_{12}-\frac{1}{9}\Phi_{22}+\frac{2}{9} \nonumber \\ 
ycc &=& -\frac{2}{3}c_a-\frac{2}{3}c_b+\frac{1}{6}\Phi_{11}-\frac{\sqrt{3}}{9}\Phi_{12}+\frac{1}{18}\Phi_{22}+\frac{5}{9}, \nonumber \\
\quad
\end{eqnarray}
by using the following conditions:
\begin{eqnarray}
c_a &=& yaa + yab + yac  \nonumber \\
c_b &=& yab + ybb + ybc  \nonumber \\
c_c &=& yac + ybc + ycc  \nonumber \\
1 &=& yaa+ybb+ycc \nonumber \\
&&+2\left(yab+yac+ybc\right).
\end{eqnarray}
%\end{widetext}
Note that above definitions of pair probability distinguish, e.g., $yab$ and $yba$ for A-B pair, relating to conventional probability $Y_{IJ}$ given by 
\begin{eqnarray}
Y_{JJ} &=& y_{JJ} \nonumber \\
Y_{IJ} &=& 2\cdot y_{IJ}\quad \left(I\neq J\right).
\end{eqnarray}
Therefore, in order to obtain pair probabilities, we should first clarify relationships between the probabilities and basis functions for multicomponent systems, which has not been explicitly provided for $R\ge 4$ systems.

%\item
We first derive the relationships for quaternary system, $R=4$. For quaternary system, orthonormal basis functions for a single lattice point are given by
\begin{eqnarray}
\phi_0 &=& 1, \quad \phi_1 = \frac{2}{\sqrt{10}}\sigma \nonumber \\
\phi_2 &=& -\frac{5}{3} + \frac{2}{3}\sigma^2,\quad \phi_3 = -\frac{17\sqrt{10}}{30}\sigma + \frac{\sqrt{10}}{6}\sigma^3
\end{eqnarray}
with $\sigma=+2,+1,-1,-2$ for element $a$, $b$, $c$, and $d$, respectively. 
In a similar fashion to ternary system, we obtain pair probabilities for, e.g., $yaa$ and $yab$ as 
\begin{widetext}
\begin{eqnarray}
\label{eq:4y}
yaa &=& \frac{1}{2}c_a+\frac{1}{10}\Phi_{11}+\frac{\sqrt{10}}{20}\Phi_{12}+\frac{1}{10}\Phi_{13}+\frac{1}{16}\Phi_{22}+\frac{\sqrt{10}}{40}\Phi_{23}+\frac{1}{40}\Phi_{33}-\frac{1}{16} \nonumber \\
yab &=& \frac{1}{4}c_a+\frac{1}{4}c_b+\frac{1}{20}\Phi_{11}-\frac{\sqrt{10}}{80}\Phi_{12}-\frac{3}{40}\Phi_{13}-\frac{1}{16}\Phi_{22}-\frac{3\sqrt{10}}{80}\Phi_{23}-\frac{1}{20}\Phi_{33}-\frac{1}{16} \nonumber \\
\end{eqnarray}
\end{widetext}

For quinary system with conventional definition of $\sigma=+2,+1,0,-1,-2$ for element A, B, C, D and E, orthonormal basis functions at a single lattice point are given by
\begin{eqnarray}
&& \phi_0 = 1,\quad \phi_1 = \frac{\sqrt{2}}{2}\sigma,\quad \phi_2 = -\sqrt{\frac{10}{7}}+\sqrt{\frac{5}{14}}\sigma^2 \nonumber \\
&& \phi_3 = -\frac{17}{6\sqrt{2}}\sigma + \frac{5}{6\sqrt{2}}\sigma^3 \nonumber \\
&& \phi_4 = \frac{3\sqrt{14}}{7} - \frac{155\sqrt{14}}{168}\sigma^2 + \frac{5\sqrt{14}}{24}\sigma^4.
\end{eqnarray}
With these basis functions, pair probabilities are given by
\begin{widetext}
\begin{eqnarray}
\label{eq:5y}
yaa &=& \frac{2}{5}c_a+\frac{2}{25}\Phi_{11}+\frac{4\sqrt{35}}{175}\Phi_{12}+\frac{2}{25}\Phi_{13}+\frac{2\sqrt{7}}{175}\Phi_{14}+\frac{2}{35}\Phi_{22}+\frac{2\sqrt{35}}{175}\Phi_{23}+\frac{2\sqrt{5}}{175}\Phi_{24}+\frac{1}{50}\Phi_{33}+\frac{\sqrt{7}}{175}\Phi_{34}+\frac{1}{350}\Phi_{44}-\frac{1}{25} \nonumber \\ 
yab &=& \frac{1}{5}c_a+\frac{1}{5}c_b+\frac{1}{25}\Phi_{11}-\frac{3}{50}\Phi_{13}-\frac{\sqrt{7}}{50}\Phi_{14}-\frac{1}{35}\Phi_{22}-\frac{\sqrt{35}}{70}\Phi_{23}-\frac{9\sqrt{5}}{350}\Phi_{24}-\frac{1}{25}\Phi_{33}-\frac{3\sqrt{7}}{175}\Phi_{34}-\frac{2}{175}\Phi_{44}-\frac{1}{25}.
\end{eqnarray}
\end{widetext}
Details for the derivation and other pair probabilities for quaternary and quinary systems are provided in Appendix.

\subsection{ Modified formulation of pair probability by a single state }
Our previous study reveal that expectation value of macroscopic structure along chosen coordination $r$ ($r$ includes both figure type and a set of basis function index) is universally given by
\begin{eqnarray}
\label{eq:emrs}
\Braket{\Phi_{r}}_Z\left(T\right) \simeq \Braket{\Phi_{r}}_{1} - \sqrt{\frac{\pi}{2}}\Braket{\Phi_{r}}_{2} \cdot \frac{U_{r}^{\textrm{proj}}}{k_{\textrm{B}}T},
\end{eqnarray}
where $\Braket{\quad}_Z$ denotes canonical average, and $\Braket{\quad}_1$ and $\Braket{\quad}_2$ respectively denotes taking arithmetic average and standard deviation over all possible microscopic states on configuration space. $U_{r}^{\rm{proj}}$ represents potential energy of the PS, whose microscopic structure is given by $\left\{\Braket{q_{1}}_{r}^{\left(+\right)}, \Braket{q_{2}}_{r}^{\left(+\right)},\ldots,\Braket{q_{g}}_{r}^{\left(+\right)}\right\}$ ($\Braket{a}_{r}^{\left(+\right)}$ denotes a partial average of scalar quantity $a$ over all microscopic states, whose structure satisfying $\Phi _{r}\ge \Braket{\Phi_r}_1$). 
While Eq.~(\ref{eq:emrs}) can be applied for any coordination under orthonormal basis, the problem is that pair probabilities discussed above are neither linear-independent nor orthonormal.
From Eq.~(\ref{eq:4p})-(\ref{eq:5y}), it is clear that each basis function always contains multiple pair probability, whose coefficient contains more than two positive or negative sign. These facts indicate that when we predict whether a \textit{single} like- and/or unlike-atom pair including a chosen element is energetically preferred in thermodynamically equilibrium state from conventional basis functions, we require energy for \textit{multiple} PS for multicomponent system leading to practically accumulating predictive error of expectation value, which is essentially different from the case of binary system. 

%\begin{itemize}
%\item
The straightforward idea to qualitatively avoid the above problems appears to starting from the use of non-conventional basis functions with appropriate definition of their domains so that resultant basis functions contain a single positive or negative coefficient for the selected pair probability. 
%\item
Let us first construct basis functions for whole lattice points by simply taking tensor product for vector space on each lattice point with basis functions, $V=\mathrm{span}\left\{1, \sigma_i , \ldots, \sigma_i^{\left(R-1\right)}\right\}$ for $R$ component system. Then we can always obtain the first two basis functions of
\begin{eqnarray}
\Phi_{11} &=& \Braket{\sigma_i \sigma_j} \nonumber \\
\Phi_{12} &=& \Braket{\sigma_i\sigma_j^2 + \sigma_i^2\sigma_j}.
\end{eqnarray}
For these functions, we can easily find that with spin variable $\sigma=-1, x_1,x_2,\cdots,x_{\left( R-1 \right)}$ ($\sigma=-1$ for a chosen element A), basis function $\Phi_{11}$ always has contain a single positive coefficient for like-atom pair A-A, and $\Phi_{12}$ always has contain negative coefficients only for unlike-atom pair including A, by satisfying the following conditions:
\begin{eqnarray}
0 < x_1,x_2,\cdots, x_{\left( R-1 \right)} < 1.
\end{eqnarray}
This leads to the fact that decrease of $\Phi_{11}$ from center of gravity (COG) of its configurational DOS (CDOS) corresponds to prefer unlike-atom pair including A, and decrease of $\Phi_{12}$ from COG of CDOS to prefer like-atom pair, A-A. However, resultant basis functions (i) are not orthogonal, and (ii) always contain multiple pair probability $y_{ij}$, it is still difficult to quantitatively determine the value of $y_{ij}$ from above single (or two) basis functions.
%\end{itemize}

%Another possible way to essentially avoid problems firstly appears to directly use probabilities $\left\{y_{ij}\right\}$, but as described, it is clear that a set of all possible probability is linear-dependent and non-orthogonal.
%, which becomes non-trivial to figure out linearly-independent set of probability when dimension and number of considered figure increases for multicomponent system, leading to making it difficult to practically find minimal condition for microscopic structure of PS. This minimal condition is practically important when numerically finding PS by minimizing distance between ideal PS and simulated PS under appropriate metric, to avoid undesired multiple counting of distances for linearly-dependent set of coordination. 
%Meanwhile, the conventional basis functions here are of course guaranteed in linear independence and are orthonormal. 
With these considerations, our strategy to essentially solve the problems, is to focusing on \textit{two} cluster  probability basis of like- and unlike-atom pair (such as $yaa$ and $yab$ with $a$ is a chosen element), that are used to determine condition of microscopic structure of PS, whose structure is \textit{not} described by cluster probability basis but by conventional basis functions. 
In order to achieve this condition, we first prepare orthonormal basis functions including a single considered pair probability $\Psi_1=y_{ij}/\sqrt{\Braket{y_{ij}|y_{ij}}}$, i.e., $\left\{\Psi_1,\cdots, \Psi_f\right\}$, where we can take any set of $\Psi_n$ ($2\le n\le f$) satisfying the orthonormality. Then we can provide
\begin{widetext}
\begin{eqnarray}
\label{eq:yij}
\Braket{y_{ij}}_Z\left(T\right) = \sqrt{\Braket{y_{ij}|y_{ij}}}\Braket{\Psi_1}_Z\left(T\right) &\simeq& \sqrt{\Braket{y_{ij}|y_{ij}}}\cdot \left(   \Braket{\Psi_1}_{1} - \sqrt{\frac{\pi}{2}} \cdot \frac{\Braket{\Psi_1}_{2}}{k_{\textrm{B}}T}\sum_{\mu}\Braket{U|\Psi_{\mu}}\Braket{\Psi_{\mu}}_{\Psi_1}^{\left(+\right)}    \right) \nonumber \\
&=&\Braket{y_{ij}}_{1} - \sqrt{\frac{\pi}{2}} \cdot \frac{\Braket{y_{ij}}_{2}}{k_{\textrm{B}}T}\sum_{\mu}\Braket{U|\Psi_{\mu}}\Braket{\Psi_{\mu}}_{y_{ij}}^{\left(+\right)} \nonumber \\
&=& \Braket{y_{ij}}_{1} - \sqrt{\frac{\pi}{2}} \cdot \frac{\Braket{y_{ij}}_{2}}{k_{\textrm{B}}T}\Braket{ \sum_\mu \sum_M \Braket{U|\Psi_\mu}\Ket{\Phi_M}\Braket{\Phi_M|\Psi_\mu}   }_{y_{ij}}^{\left(+\right)} \nonumber \\
&=& \Braket{y_{ij}}_{1} - \sqrt{\frac{\pi}{2}} \cdot \frac{\Braket{y_{ij}}_{2}}{k_{\textrm{B}}T}\sum_{M}\Braket{U|\Phi_{M}}\Braket{\Phi_{M}}_{y_{ij}}^{\left(+\right)},
\quad 
\end{eqnarray}
\end{widetext}
where conventional pair probability $Y_{IJ}$ is given by
\begin{eqnarray}
\Braket{Y_{IJ}}_Z\left(T\right) = \left(2 - \delta_{IJ}\right)\cdot \Braket{y_{IJ}}_Z\left(T\right).
\end{eqnarray}
To obtain the last equation, we employ linearity for average and standard deviation of $\Braket{\quad}_1$ and $\Braket{\quad}_2$, completeness of $\sum_\mu \Ket{\Psi_\mu}\Bra{\Psi_\mu}=\sum_M\Ket{\Phi_M}\Bra{\Phi_M}=\mathbf{1}$, and inner products (trace over configuration space) does not by definition depend on partial average of $\Braket{\quad}_{y_{ij}}^{\left(+\right)}$.
In Eq.~(\ref{eq:yij}), corresponding projection state for pair probability $yij$ is described in terms of $\Braket{\Phi_{M}}_{y_{ij}}^{\left(+\right)}$ ($M$ denotes all possible combination of basis index and figure), which can be practically obtained from Eq.~(\ref{eq:4y}) and Eq.~(\ref{eq:5y}) for quaternary and quinary system, respectively. 
%This corresponds to applying non-orthogonal basis including $yaa$, $yab$ and rest of linearly-independent basis functions to Eq.~(\ref{eq:emrs}), where microscopic structure is considered by conventional basis functions. 
The great advantage of the present approach is thus (i) pair probability can be determined from a single PS with structure of $\left\{ \Braket{\Phi_{1}}_{y_{ij}}^{\left(+\right)}, \cdots, \Braket{\Phi_{f}}_{y_{ij}}^{\left(+\right)} \right\}$ , and (ii) without requiring explicit expression of orthonormal basis including desired pair probabilities (i.e., without knowing $\left\{\Psi_2,\cdots, \Psi_f\right\}$), orthonormality is exactly guaranteed by conventional basis functions to construct PS. 

\subsection{Application to Multicomponent System}
\begin{figure}[h]
\begin{center}
\includegraphics[width=0.95\linewidth]{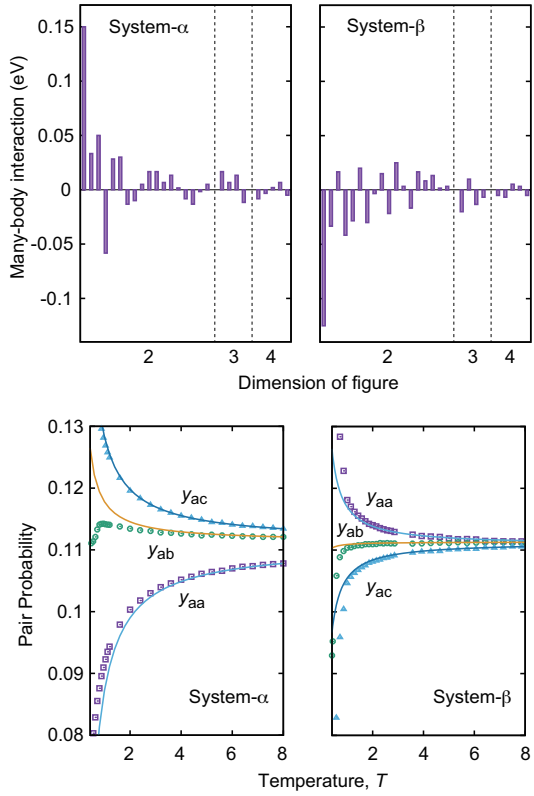}
\caption{Upper: Many-body interactions for two artificially-prepared ternary systems $\alpha$ and $\beta$. Lower: Temperature dependence of pair probabilities obtained by conventional thermodynamic simulation (open triangles, circles and squares) and by the present approach based on the PS energy. }
\label{fig:sro}
\end{center}
\end{figure}
We finally demonstrate the validity of the modified formulation for pair probability, Eq.~(\ref{eq:yij}). 
We artificially prepare two sets of many-body interaction consisting of up to 6-th neighbor (6NN) pair, and minimal triplet and quartet figures on ternary fcc lattice, resulting in totally 27 interactions as shown in upper side of Fig.~\ref{fig:sro}. Using these interactions, we employ Monte Carlo (MC) statistical simulation under canonical ensemble to estimate temperature dependence of pair probabilities of $yaa$, $yab$ and $yac$ based on the MC cell with 4800 atoms (i.e., $12\times10\times10$ expansion of conventional unit cell) and 10000 MC steps per site. In order to obtain the values of $\Braket{\Phi_{M}}_{y_{ij}}^{\left(+\right)}$ for two projection states along $yaa$ and $yab$, we also perform MC simulation to uniformly sampling microscopic states on configuration space with the same calculation conditions as above. The structural information for the two PS is then applied to Eq.~(\ref{eq:yij}) to obtain $T$-dependence of pair probabilities by the present approach. The results are summarized in the lower-part of Fig.~\ref{fig:sro}. For conventional thermodynamic simulation, we can clearly see that system $\alpha$ and $\beta$ exhibit completely different ordering tendency: System $\alpha$ exhibits unlike-atom pair preference, while system $\beta$ prefers like-atom pair, i.e., undergoing to phase separation. Such ordering tendency can be reasonably characterized by the present approach of Eq.~(\ref{eq:yij}), where again, the structural information about the single PS along chosen pair is \textit{common} for system $\alpha$ and $\beta$. Thus we can see the validity of the proposed approach to predicting selected pair probability for multicomponent system based on information about a single specially selected microscopic state.

\section{Conclusions}
We here show that by deriving complete relationships between conventional basis functions and pair probability, expectation value of like- and unlike-atom pair including a chosen single element in equilibrium state for multicomponent system can be directly determined from a single projection state (PS). Condition for microscopic structure of the PS is also provided. These can avoid use of multiple PS energy to predict a single pair probability, leading to avoid accumulation of predictive error of energy coming from individual PS. 

\section{Acknowledgement}
This work is supported by a Grant-in-Aid for Scientific Research on Innovative Areas (18H05453) and a Grant-in-Aid for Scientific Research (16K06704) from the MEXT of Japan, Research Grant from Hitachi Metals$\cdot$Materials Science Foundation, and Advanced Low Carbon Technology Research and Development Program of the Japan Science and Technology Agency (JST).

\section*{Appendix}

\appendix

\section*{Derivation of relationships between basis function and pair probability for $R=4,5$.}
In a similar fashion to ternary system of $R=3$, basis functions for a chosen pair $\alpha$ for e.g., combination of basis function $\phi^{\left(1\right)}$ and $\phi^{\left(1\right)}$ can be written by constituent pair probability $y{ij}$ ($i,j=a,b,c,d$):
\begin{widetext}
\begin{eqnarray}
\label{eq:40}
\Phi_{11} = \left(\frac{2}{\sqrt{10}}\right)^2 \sum_{i,j} y_{ij}\cdot \sigma_i\sigma_j = \frac{2}{5}\left(4yaa+4yab-4yac-8yad+ybb-2ybc-4ybd+ycc+4ycd+4ydd\right).
\end{eqnarray}
\end{widetext}
In a similar fashion, other basis functions can also be given by
\begin{widetext}
\begin{eqnarray}
\label{eq:4p}
\Phi_{12} &=& \frac{2}{3\sqrt{10}}\left(6yaa-3yab-9yac-3ybb+9ybd+3ycc+3ycd-6ydd\right) \nonumber \\
\Phi_{13} &=& \frac{1}{5}\left(4yaa-6yab+6yac-8yad-4ybb+8ybc+6ybd-4ycc-6ycd+4ydd\right) \nonumber \\
\Phi_{22} &=& yaa-2yab-2yac+2yad+ybb+2ybc-2ybd+ycc-2ycd+ydd \nonumber \\
\Phi_{23} &=& \frac{\sqrt{10}}{5} \left(yaa-3yab+yac+2ybb-ybd-2ycc+3ycd-ydd\right) \nonumber \\
\Phi_{33} &=& \frac{1}{5}\left(2yaa-8yab+8yac-4yad+8ybb-16ybc+8ybd+8ycc-8ycd+2ydd\right).
\end{eqnarray}
\end{widetext}
Using Eqs.~(\ref{eq:40}) and (\ref{eq:4p}), and the following conditions for pair probability and composition $c_i$ for element $i$
\begin{eqnarray}
c_a &=& yaa + yab + yac + yad \nonumber \\
c_b &=& yab + ybb + ybc + ybd \nonumber \\
c_c &=& yac + ybc + ycc + ycd \nonumber \\
1 &=& yaa+ybb+ycc+ydd \nonumber \\
&&+2\left(yab+yac+yad+ybc+ybd+ycd\right),
\end{eqnarray}
we can explicitly describe pair probabilities in terms of basis functions:
\begin{widetext}
\begin{eqnarray}
\label{eq:4y}
yaa &=& \frac{1}{2}c_a+\frac{1}{10}\Phi_{11}+\frac{\sqrt{10}}{20}\Phi_{12}+\frac{1}{10}\Phi_{13}+\frac{1}{16}\Phi_{22}+\frac{\sqrt{10}}{40}\Phi_{23}+\frac{1}{40}\Phi_{33}-\frac{1}{16} \nonumber \\
yab &=& \frac{1}{4}c_a+\frac{1}{4}c_b+\frac{1}{20}\Phi_{11}-\frac{\sqrt{10}}{80}\Phi_{12}-\frac{3}{40}\Phi_{13}-\frac{1}{16}\Phi_{22}-\frac{3\sqrt{10}}{80}\Phi_{23}-\frac{1}{20}\Phi_{33}-\frac{1}{16} \nonumber \\
yac &=& \frac{1}{4}c_a+\frac{1}{4}c_c-\frac{1}{20}\Phi_{11}-\frac{3\sqrt{10}}{80}\Phi_{12}+\frac{3}{40}\Phi_{13}-\frac{1}{16}\Phi_{22}+\frac{\sqrt{10}}{80}\Phi_{23}+\frac{1}{20}\Phi_{33}-\frac{1}{16} \nonumber \\ 
yad &=& -\frac{1}{4}c_b-\frac{1}{4}c_c-\frac{1}{10}\Phi_{11}-\frac{1}{10}\Phi_{13}+\frac{1}{16}\Phi_{22}-\frac{1}{40}\Phi_{33}+\frac{3}{16} \nonumber \\
ybb &=& \frac{1}{2}c_b+\frac{1}{40}\Phi_{11}-\frac{\sqrt{10}}{40}\Phi_{12}-\frac{1}{10}\Phi_{13}+\frac{1}{16}\Phi_{22}+\frac{\sqrt{10}}{20}\Phi_{23}+\frac{1}{10}\Phi_{33}-\frac{1}{16} \nonumber \\ 
ybc &=& \frac{1}{4}c_b+\frac{1}{4}c_c-\frac{1}{40}\Phi_{11}+\frac{1}{10}\Phi_{13}+\frac{1}{16}\Phi_{22}-\frac{1}{10}\Phi_{33}-\frac{1}{16} \nonumber \\
ybd &=& -\frac{1}{4}c_a-\frac{1}{4}c_c-\frac{1}{20}\Phi_{11}+\frac{3\sqrt{10}}{80}\Phi_{12}+\frac{3}{40}\Phi_{13}-\frac{1}{16}\Phi_{22}-\frac{\sqrt{10}}{80}\Phi_{23}+\frac{1}{20}\Phi_{33}+\frac{3}{16} \nonumber \\
ycc &=& \frac{1}{2}c_c+\frac{1}{40}\Phi_{11}+\frac{\sqrt{10}}{40}\Phi_{12}-\frac{1}{10}\Phi_{13}+\frac{1}{16}\Phi_{22}-\frac{\sqrt{10}}{20}\Phi_{23}+\frac{1}{10}\Phi_{33}-\frac{1}{16} \nonumber \\
ycd &=& -\frac{1}{4}c_a-\frac{1}{4}c_b+\frac{1}{20}\Phi_{11}+\frac{\sqrt{10}}{80}\Phi_{12}-\frac{3}{40}\Phi_{13}-\frac{1}{16}\Phi_{22}+\frac{3\sqrt{10}}{80}\Phi_{23}-\frac{1}{20}\Phi_{33}+\frac{3}{16} \nonumber \\
ydd &=& -\frac{1}{2}c_a-\frac{1}{2}c_b-\frac{1}{2}c_c+{\frac {{1}}{10}}\Phi_{11}-\frac{\sqrt{10}}{20}\Phi_{12}+\frac{1}{10}\Phi_{13}+\frac{1}{16}\Phi_{22}-\frac{\sqrt{10}}{40}\Phi_{23}+\frac{1}{40}\Phi_{33}+{\frac{7}{16}}.
\end{eqnarray}
\end{widetext}

For quinary system, basis functions for a chosen pair figure can be rewritten in terms of corresponding pair probabilities:
\begin{widetext}
\begin{eqnarray}
\label{eq:5p}
\Phi_{11} &=& \frac{1}{2}\left(4yaa+4yab-4yad-8yae+ybb-2ybd-4ybe+ydd+4yde+4yee\right) \nonumber \\
\Phi_{12} &=& \frac{\sqrt{35}}{14}\left(4yaa-4yad-ybb+4ybe-4yac-2ybc+2ycd+4yce+ydd-4yee\right) \nonumber \\
\Phi_{13} &=& \frac{1}{2}\left(2yaa-3yab+3yad-4yae-2ybb+4ybd+3ybe-2ydd-3yde+2yee\right) \nonumber \\
\Phi_{14} &=& \frac{\sqrt{7}}{14}\left(2yaa-7yab-9yad-4ybb+9ybe+12yac+6ybc-6ycd-12yce+4ydd+7yde-2yee\right) \nonumber \\
\Phi_{22} &=& \frac{5}{14}\left(4yee-4yde+ydd+4ycd-8yce+4ycc+2ybd-4ybe+ybb+4ybc+8yae-4yad-4yab-8yac+4yaa\right) \nonumber \\
\Phi_{23} &=& \frac{\sqrt{35}}{14}\left(2yaa-5yab-2yac+3yad+2ybb+4ybc-3ybe-4ycd-2ydd+5yde+2yce-2yee\right) \nonumber \\
\Phi_{24} &=& \frac{\sqrt{5}}{14}\left(2yee+4ydd-9yde+10yce-12ycc+2ycd-9ybe+2ybc+8ybd+4ybb+4yae+10yac-9yad-9yab+2yaa\right) \nonumber \\
\Phi_{33} &=& \frac{1}{2}\left(yaa-4yab+4yad-2yae+4ybb-8ybd+4ybe+4ydd-4yde+yee\right) \nonumber \\
\Phi_{34} &=& \frac{\sqrt{7}}{14}\left(yaa-6yab-2yad+8ybb+2ybe+6yac-12ybc+12ycd-6yce-8ydd+6yde-yee\right) \nonumber \\
\Phi_{44} &=& \frac{1}{14}\left(yaa-8yab+12yac-8yad+2yae+16ybb-48ybc+32ybd-8ybe+36ycc-48ycd+12yce+16ydd-8yde+yee\right). \nonumber \\
\quad
\end{eqnarray}
\end{widetext}
Using Eq.~(\ref{eq:5p}) and the following conditions for pair probability and composition $c_i$ for element $i$
\begin{eqnarray}
c_a &=& yaa + yab + yac + yad + yae \nonumber \\
c_b &=& yab + ybb + ybc + ybd +ybe\nonumber \\
c_c &=& yac + ybc + ycc + ycd + yce\nonumber \\
c_d &=& yad + ybd + ycd + ydd + yde \nonumber \\
1 &=& yaa+ybb+ycc+ydd+yee \nonumber \\
&+&2\left(yab+yac+yad+yae+ybc+\right. \nonumber \\
&&\left. ybd+ybe+ycd+yce+yde\right),
\end{eqnarray}
we can explicitly describe pair probabilities for quinary system: 
\begin{widetext}
\begin{eqnarray}
\label{eq:5y}
yaa &=& \frac{2}{5}c_a+\frac{2}{25}\Phi_{11}+\frac{4\sqrt{35}}{175}\Phi_{12}+\frac{2}{25}\Phi_{13}+\frac{2\sqrt{7}}{175}\Phi_{14}+\frac{2}{35}\Phi_{22}+\frac{2\sqrt{35}}{175}\Phi_{23}+\frac{2\sqrt{5}}{175}\Phi_{24}+\frac{1}{50}\Phi_{33}+\frac{\sqrt{7}}{175}\Phi_{34}+\frac{1}{350}\Phi_{44}-\frac{1}{25} \nonumber \\ 
yab &=& \frac{1}{5}c_a+\frac{1}{5}c_b+\frac{1}{25}\Phi_{11}-\frac{3}{50}\Phi_{13}-\frac{\sqrt{7}}{50}\Phi_{14}-\frac{1}{35}\Phi_{22}-\frac{\sqrt{35}}{70}\Phi_{23}-\frac{9\sqrt{5}}{350}\Phi_{24}-\frac{1}{25}\Phi_{33}-\frac{3\sqrt{7}}{175}\Phi_{34}-\frac{2}{175}\Phi_{44}-\frac{1}{25} \nonumber \\
yac &=& \frac{1}{5}c_a+\frac{1}{5}c_c-\frac{2\sqrt{35}}{175}\Phi_{12}+\frac{6\sqrt{7}}{175}\Phi_{14}-\frac{2}{35}\Phi_{22}-\frac{\sqrt{35}}{175}\Phi_{23}+\frac{\sqrt{5}}{35}\Phi_{24}+\frac{3\sqrt{7}}{175}\Phi_{34}+\frac{3}{175}\Phi_{44}-\frac{1}{25} \nonumber \\
yad &=& \frac{1}{5}c_a+\frac{1}{5}c_d-\frac{1}{25}\Phi_{11}-\frac{2\sqrt{35}}{175}\Phi_{12}+\frac{3}{50}\Phi_{13}-\frac{9\sqrt{7}}{350}\Phi_{14}-\frac{1}{35}\Phi_{22}+\frac{3\sqrt{35}}{350}\Phi_{23}-\frac{9\sqrt{5}}{350}\Phi_{24}+\frac{1}{25}\Phi_{33}-\frac{\sqrt{7}}{175}\Phi_{34}-\frac{2}{175}\Phi_{44} \nonumber \\
&&-\frac{1}{25} \nonumber \\
yae &=& -\frac{1}{5}c_b-\frac{1}{5}c_c-\frac{1}{5}c_d-\frac{2}{25}\Phi_{11}-\frac{2}{25}\Phi_{13}+\frac{2}{35}\Phi_{22}+\frac{2\sqrt{5}}{175}\Phi_{24}-\frac{1}{50}\Phi_{33}+\frac{1}{350}\Phi_{44}+\frac{4}{25} \nonumber \\ 
ybb &=& \frac{2}{5}c_b+\frac{1}{50}\Phi_{11}-\frac{\sqrt{35}}{175}\Phi_{12}-\frac{2}{25}\Phi_{13}-\frac{4\sqrt{7}}{175}\Phi_{14}+\frac{1}{70}\Phi_{22}+\frac{2\sqrt{35}}{175}\Phi_{23}+\frac{4\sqrt{5}}{175}\Phi_{24}+\frac{2}{25}\Phi_{33}+\frac{8\sqrt{7}}{175}\Phi_{34}+\frac{8}{175}\Phi_{44}-\frac{1}{25} \nonumber \\
ybc &=& \frac{1}{5}c_b+\frac{1}{5}c_c-\frac{\sqrt{35}}{175}\Phi_{12}+\frac{3\sqrt{7}}{175}\Phi_{14}+\frac{1}{35}\Phi_{22}+\frac{2\sqrt{35}}{175}\Phi_{23}+\frac{\sqrt{5}}{175}\Phi_{24}-\frac{6\sqrt{7}}{175}\Phi_{34}-\frac{12}{175}\Phi_{44}-\frac{1}{25} \nonumber \\
ybd &=& \frac{1}{5}c_b+\frac{1}{5}c_d-\frac{1}{50}\Phi_{11}+\frac{2}{25}\Phi_{13}+\frac{1}{70}\Phi_{22}+\frac{4\sqrt{5}}{175}\Phi_{24}-\frac{2}{25}\Phi_{33}+\frac{8}{175}\Phi_{44}-\frac{1}{25} \nonumber \\
ybe &=& -\frac{1}{5}c_a-\frac{1}{5}c_c-\frac{1}{5}c_d-\frac{1}{25}\Phi_{11}+\frac{2\sqrt{35}}{175}\Phi_{12}+\frac{3}{50}\Phi_{13}+\frac{9\sqrt{7}}{350}\Phi_{14}-\frac{1}{35}\Phi_{22}-\frac{3\sqrt{35}}{350}\Phi_{23}-\frac{9\sqrt{5}}{350}\Phi_{24}+\frac{1}{25}\Phi_{33}+\frac{\sqrt{7}}{175}\Phi_{34}\nonumber \\
&&-\frac{2}{175}\Phi_{44}+\frac{4}{25} \nonumber \\
ycc &=& \frac{2}{5}c_c+\frac{2}{35}\Phi_{22}-\frac{12\sqrt{5}}{175}\Phi_{24}+\frac{18}{175}\Phi_{44}-\frac{1}{25} \nonumber \\
ycd &=& \frac{1}{5}c_c+\frac{1}{5}c_d+\frac{\sqrt{35}}{175}\Phi_{12}-\frac{3\sqrt{7}}{175}\Phi_{14}+\frac{1}{35}\Phi_{22}-\frac{2\sqrt{35}}{175}\Phi_{23}+\frac{\sqrt{5}}{175}\Phi_{24}+\frac{6\sqrt{7}}{175}\Phi_{34}-\frac{12}{175}\Phi_{44}-\frac{1}{25} \nonumber \\
 yce &=& -\frac{1}{5}c_a-\frac{1}{5}c_b-\frac{1}{5}c_d+\frac{2\sqrt{35}}{175}\Phi_{12}-\frac{6\sqrt{7}}{175}\Phi_{14}-\frac{2}{35}\Phi_{22}+\frac{\sqrt{35}}{175}\Phi_{23}+\frac{\sqrt{5}}{35}\Phi_{24}-\frac{3\sqrt{7}}{175}\Phi_{34}+\frac{3}{175}\Phi_{44}+\frac{4}{25} \nonumber \\
ydd &=& \frac{2}{5}c_d+\frac{1}{50}\Phi_{11}+\frac{\sqrt{35}}{175}\Phi_{12}-\frac{2}{25}\Phi_{13}+\frac{4\sqrt{7}}{175}\Phi_{14}+\frac{1}{70}\Phi_{22}-\frac{2\sqrt{35}}{175}\Phi_{23}+\frac{4\sqrt{5}}{175}\Phi_{24}+\frac{2}{25}\Phi_{33}-\frac{8\sqrt{7}}{175}\Phi_{34}+\frac{8}{175}\Phi_{44}-\frac{1}{25} \nonumber \\
yde &=& -\frac{1}{5}c_a-\frac{1}{5}c_b-\frac{1}{5}c_c+\frac{1}{25}\Phi_{11}-\frac{3}{50}\Phi_{13}+\frac{\sqrt{7}}{50}\Phi_{14}-\frac{1}{35}\Phi_{22}+\frac{\sqrt{35}}{70}\Phi_{23}-\frac{9\sqrt{5}}{350}\Phi_{24}-\frac{1}{25}\Phi_{33}+\frac{3\sqrt{7}}{175}\Phi_{34}-\frac{2}{175}\Phi_{44}+\frac{4}{25} \nonumber \\
yee &=& -\frac{2}{5}c_a-\frac{2}{5}c_b-\frac{2}{5}c_c-\frac{2}{5}c_d+\frac{2}{25}\Phi_{11}-\frac{4\sqrt{35}}{175}\Phi_{12}+\frac{2}{25}\Phi_{13}-\frac{2\sqrt{7}}{175}\Phi_{14}+\frac{2}{35}\Phi_{22}-\frac{2\sqrt{35}}{175}\Phi_{23}+\frac{2\sqrt{5}}{175}\Phi_{24}+\frac{1}{50}\Phi_{33}\nonumber \\
&&-\frac{\sqrt{7}}{175}\Phi_{34}+\frac{1}{350}\Phi_{44}+\frac{9}{25} \nonumber \\
\end{eqnarray}
\end{widetext}

\end{document}